\documentclass[prb,preprint,superscriptaddress,nobibnotes]{revtex4}

\usepackage{graphicx}
\usepackage{epstopdf}
\usepackage{dcolumn}
\usepackage{bm}
\usepackage{times,mathptmx}
\usepackage{color}
\usepackage{multirow}
\usepackage{enumitem}
\usepackage{chngpage}
\usepackage[T1]{fontenc}

\begin{document}
\title{\bf Polygon Coherent Modes in a Weakly Perturbed Whispering Gallery Microresonator for Efficient Second Harmonic, Optomechanic and Frequency Comb Generations}

\author{Zhiwei Fang}
\thanks{These authors contributed equally to this work.}
\affiliation{State Key Laboratory of Precision Spectroscopy, East China Normal University, Shanghai 200062, China.}
\affiliation{XXL{-}The Extreme Optoelectromechanics Laboratory, School of Physics and Electric Science, East China Normal University, Shanghai 200241, China.}

\author{Sanaul Haque}
\thanks{These authors contributed equally to this work.}
\affiliation{Department of Electrical and Computer Engineering, University of Victoria, Victoria, BC V8P 5C2, Canada}

\author{Farajollahi Saeed}
\affiliation{Department of Electrical and Computer Engineering, University of Victoria, Victoria, BC V8P 5C2, Canada}

\author{Haipeng Luo}
\affiliation{Department of Electrical and Computer Engineering, University of Victoria, Victoria, BC V8P 5C2, Canada}

\author{Jintian Lin}
\affiliation{State Key Laboratory of High Field Laser Physics, Shanghai Institute of Optics and Fine Mechanics, Chinese Academy of Sciences, Shanghai 201800, China.}

\author{Rongbo Wu}
\affiliation{State Key Laboratory of High Field Laser Physics, Shanghai Institute of Optics and Fine Mechanics, Chinese Academy of Sciences, Shanghai 201800, China.}

\author{Jianhao Zhang}
\affiliation{State Key Laboratory of High Field Laser Physics, Shanghai Institute of Optics and Fine Mechanics, Chinese Academy of Sciences, Shanghai 201800, China.}

\author{Zhe Wang}
\affiliation{State Key Laboratory of High Field Laser Physics, Shanghai Institute of Optics and Fine Mechanics, Chinese Academy of Sciences, Shanghai 201800, China.}

\author{Min Wang}
\affiliation{State Key Laboratory of Precision Spectroscopy, East China Normal University, Shanghai 200062, China.}
\affiliation{XXL{-}The Extreme Optoelectromechanics Laboratory, School of Physics and Electric Science, East China Normal University, Shanghai 200241, China.}

\author{Ya Cheng}
\email{ya.cheng@siom.ac.cn}
\affiliation{State Key Laboratory of Precision Spectroscopy, East China Normal University, Shanghai 200062, China.}
\affiliation{XXL{-}The Extreme Optoelectromechanics Laboratory, School of Physics and Electric Science, East China Normal University, Shanghai 200241, China.}
\affiliation{State Key Laboratory of High Field Laser Physics, Shanghai Institute of Optics and Fine Mechanics, Chinese Academy of Sciences, Shanghai 201800, China.}
\affiliation{Collaborative Innovation Center of Extreme Optics, Shanxi University, Taiyuan 030006, Shanxi, China.}

\author{Tao Lu}
\email{taolu@ece.uvic.ca}
\affiliation{Department of Electrical and Computer Engineering, University of Victoria, Victoria, BC V8P 5C2, Canada}

\maketitle

{\bf We observe high optical quality factor (Q) polygonal and star coherent optical modes in a lithium niobate microdisk.  In contrast to the previous polygon modes achieved by deformed microcavities at lower mechanical and optical Q, we adopted weak perturbation from a tapered fiber for the polygon mode formation. The resulting high intracavity optical power of the polygon modes triggers second harmonic generation at high efficiency. With the combined advantage of high mechanical Q cavity optomechanical oscillation was observed for the first time. Finally, we observe frequency microcomb generation from the polygon modes with an ultra stable taper-on-disk coupling mechanism.}

Polygon shaped quantum wave functions in perturbed circular mesoscopic billiards are coherent combinations of eigenstates whose eigenvalues are close. The exploration of these coherent states leads us to deeper understanding on the interrelation between quantum and classical worlds~\cite{Liu_Wave}. In contrast, polygonal optical modes can be formed through coherent re-combination of conventional whispering gallery micro-resonator (WGM) optical modes provided their resonance wavelengths are close in value. This provide us an intriguing tool on viewing chaotic effects as well as achieving unidirectional lasing~\cite{Gmachl_High,Shinohara_Chaos,Song_Directional,Cao_dielectric}. In the past, polygon, semi-polygon and star optical trajectories were observed on deformed WGMs~\cite{Redding_Local,doi:10.1063/1.2535692,Unterhinninghofen_Goos,PhysRevA.72.013803,Wiersig_Combining,PhysRevLett.88.094102}. In these studies, the deformation breaks the cavity azimuthal symmetric and provide strong perturbations that enables coherent recombination of the circularly symmetry whispering gallery modes to form new resonant modes. These modes, under certain circumstances, display non-circular patterns as mentioned above. The microcavity deformation, however, sacrifices the cavity optical quality factor (Q) and consequently reduces the intracavity power that could otherwise trigger a plural of nonlinear optical effects. To circumvent this problem, here we report the observations of polygon and star modes in a high Q microcavity. In our study, the coherent recombination was achieved through weak perturbation via a tapered optical fiber with minimum impairment on optical Q. Consequently, the polygon modes may process Q as high as $10^6$.  As demonstrated in the rest of the article, the resulting large intracavity circulating power triggers frequency comb generation that have never been reported with polygon modes. Further, larger modal overlap between the fundamental and second harmonic same-polygon modes enhances high efficiency second harmonic generations compared to that observed from deformed cavities.

\section*{Results}
\subsection*{Theory of polygon modes in WGM} Ideal whispering gallery microcavities are azimuthal symmetric. Under the theory of whispering gallery modes, these modes are approximately mutually orthogonal when the optical Q is sufficiently high. In the presents of perturbations, however, the eigen modes are no longer orthogonal. Consequently, a new set of orthogonal resonant modes will be formed through the linear combination of the unperturbed modes. Since the perturbation breaks the circular symmetry, the new resonance modes can be azimuthal dependent. In particular, with sufficient number of unperturbed modes involved and if their resonance frequencies are close enough, the coherent superposition of these eigen modes may form polygon patterns. The details of perturbed cavity mode theory can be further explained through mode matching approach~\cite{Du2013a,Du:14} while the numerical modelling can be done through eg. finite element techniques accordingly. In the past, such perturbations and symmetry breaking were introduced through the deformation of cavity shape, which is inadvisable to degrade its optical Q to as low as $10^4$.~\cite{Shinohara_Chaos,Song_Directional} In addition, the mechanical Q of the cavity will also be reduced as a result of the deformation. Therefore, even on resonance, the intracavity power of the polygon modes are relatively low. Besides, the materials of the deformation of cavity are not the nonlinear optical crystal. That impairs its capability to excite nonlinear effects where sufficient optical power is essential. Therefore, the study of this article is to form polygon modes through weaker perturbation with minimum reduction of optical Q.
\begin{figure}[hbtp]
   \centering
  \includegraphics[width=0.90\textwidth]{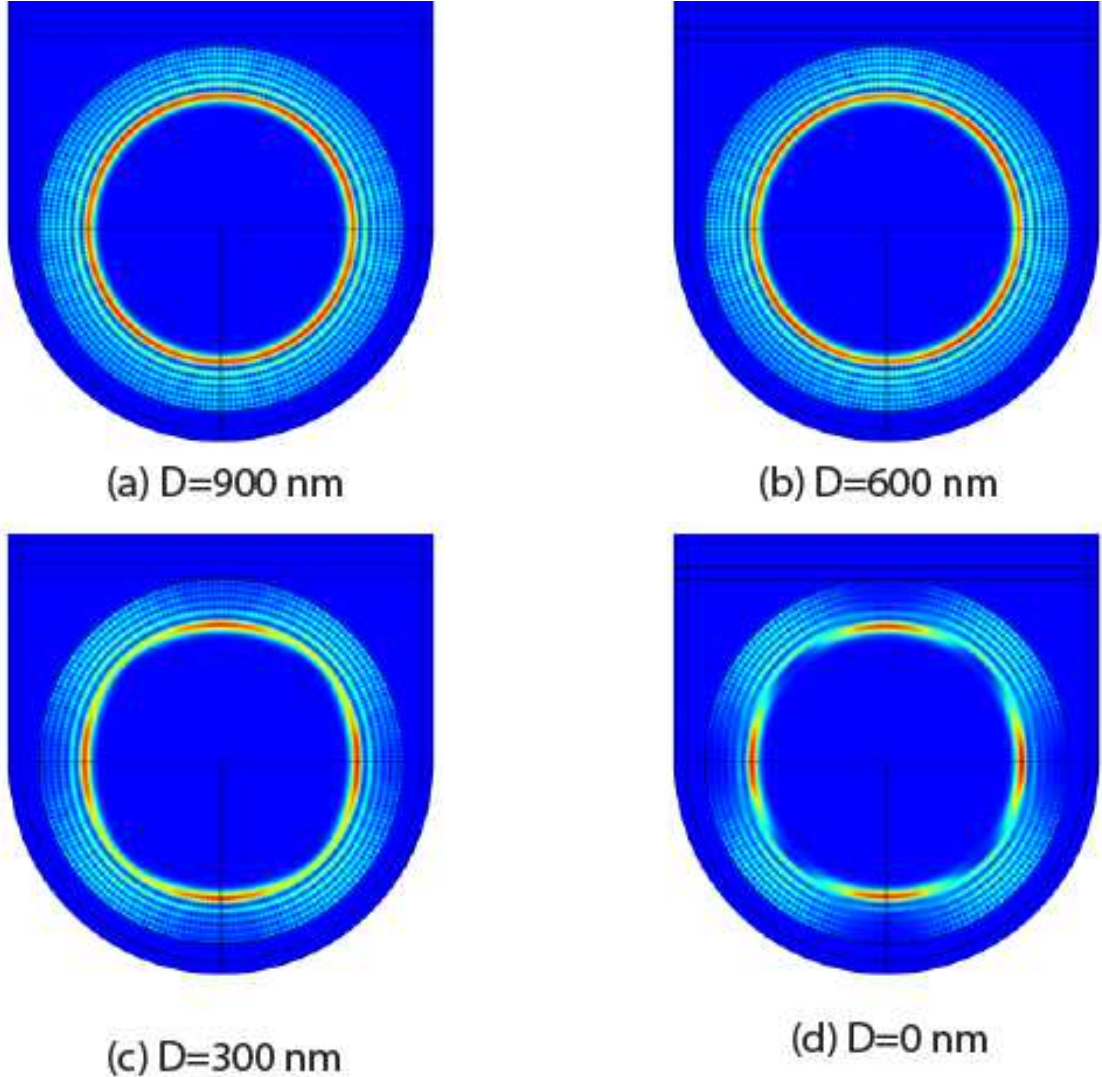}
   \caption{Gradual formation of square mode when the gap (D) between the tapered fiber and microdisk reduces from (a) 900~nm to (b) 600~nm, (c) 300~nm and (d) 0~nm.}\label{gap_evolve}
\end{figure}
To introduce weak perturbation for polygon mode formation, we simply choose the tapered fiber that was typically used as coupling element in WGM experiments. We observe that it provides the ideal amount of perturbation to form the polygon modes while through adjusting the disk/taper distance we can reduce the coupling loss to maintain the optical Q at high level. To demonstrate, we simulate the taper/microdisk system through COMSOL. As shown in Fig.~\ref{gap_evolve}, clearly when the gap between the tapered fiber and the microdisk is far (D=900~nm), the intensity distribution of the resonant mode is still circularly symmetric. When the gap reduces to 600~nm and 300~nm, periodic intensity fluctuations along the azimuthal direction becomes gradually visible. This fluctuation further evolves to a square shape when the gap reduces to 0~nm. Evidently, the introduction of the fiber taper triggers the polygon mode formation in an close-to-idea circular microdisk.

\subsection*{Direct Observation of Polygon and Star Modes} 
\begin{figure}[hbtp]
   \centering
  \includegraphics[width=0.90\textwidth]{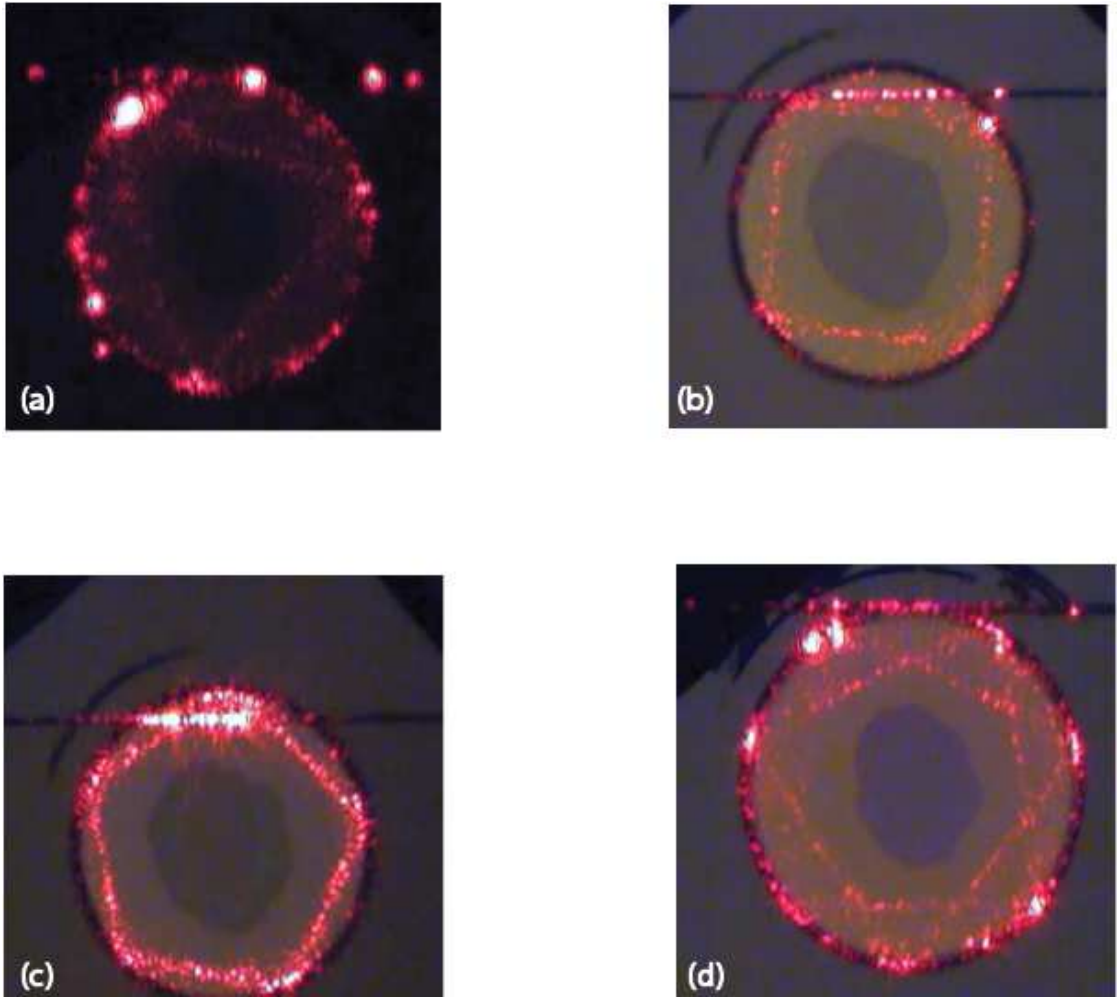}
 \caption{(a) Triangle mode, (b) Square mode, (c) Pentagon mode and (d) Star mode. All excited by 632~nm laser.}\label{passive_modes_all}
\end{figure}

In experiment, we verify the existence of the taper induced polygon modes on a high Q lithium niobate microdisk. We first use the visible 632~nm tunable laser to probe the cavity resonant modes and view the mode pattern under an optical microscope. As shown in Fig.~\ref{passive_modes_all}, at different wavelengths, triangular, square and pentagon modes were observed in subplots Fig.~\ref{passive_modes_all}a-c. It is worth mentioning that to excite a regular whispering gallery mode, the tapered fiber has to be placed carefully at the edge of the cavity, which generally causes the disk/taper gap fluctuations and destabilizes the cavity light output in addition to the extraordinary difficulty in alignment. In the case of polygon mode excitation, however, taper can be placed on the top of and in contact to the disk as shown in the subplot Fig.~\ref{passive_modes_all}b and~\ref{passive_modes_all}c. Such arrangement not only greatly facility the taper/disk alignment but also stabilize the light output by removing the gap between the disk and the tapered fiber. Further, we also observed star pattern as shown in Fig.~\ref{passive_modes_all}d, confirming that the taper fiber can fully replace the cavity deformation to trigger the polygon and star pattern formation.

\subsection*{Second harmonic generation in polygon modes} 
\begin{figure}[hbtp]
   \centering
  \includegraphics[width=0.90\textwidth]{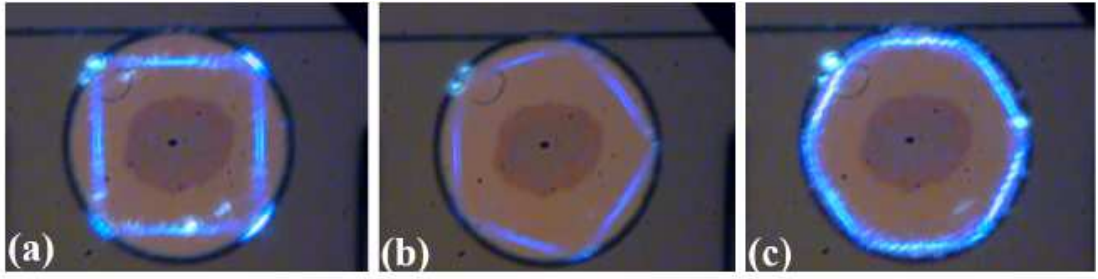}
 \caption{Second harmonic of (a) Square mode, (b) Pentagon mode, (c) Hexagon mode. All excited by a 970~nm pump laser.}\label{Fig_2nd}
\end{figure}

In the second experiment, we observed the clear polygon mode pattern in second harmonic generation (SHG) when pumped by a 970~nm wavelength laser. As shown in Fig.~\ref{Fig_2nd}, the visible SHG mode (${\sim}$485~nm) displays square, pentagon and hexagon pattern. In this experiment, the quadratic pump power dependency of the SHG signal enhanced the contrast of the polygon pattern.

\begin{figure}[hbtp]
   \centering
  \includegraphics[width=0.90\textwidth]{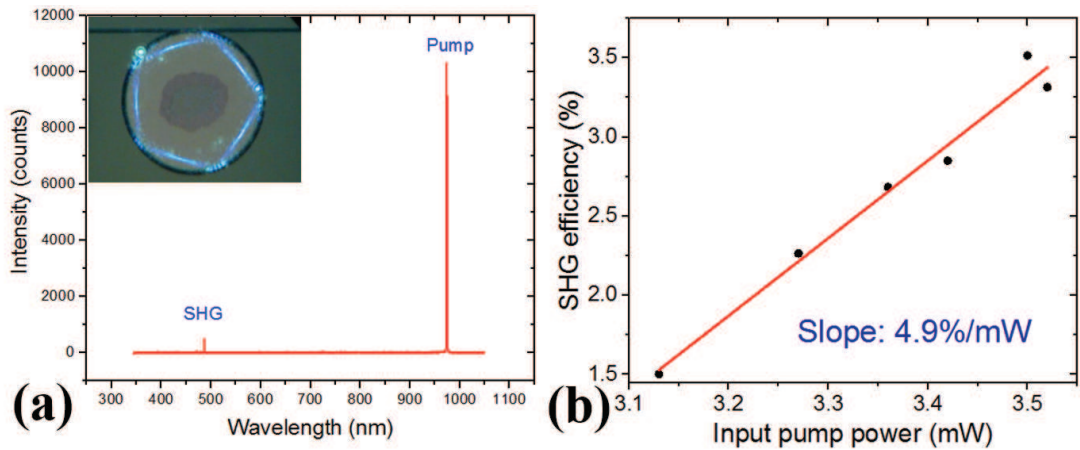}
\caption{(a) The optical spectrum of the second harmonic and its pump light. (b) The SHG conversion efficiency as a function of the pump power.}\label{second_harmonic_efficiency}
\end{figure}

In the following experiment, we further measured the SHG efficiency. During the experiment, we linearly scanned the optical frequency of the probe laser and adjusted the coupling between the tapered fiber and LN microdisk. Fig.~\ref{second_harmonic_efficiency}a shows the optical spectrum in a pentagon mode at a wavelength of 486.7~nm (top left inset). The wavelength of the pump light is 973.4~nm, which is exactly twice of that of the second harmonic signal. We notice that the second harmonic signal can be seen even when the pump signal is far from saturation, indicating the conversion efficiency is very high. To measure the conversion efficiency, we recording the power of the pump $(P_{Pump})$ and the spectrogram synchronously and obtain $\frac{P_{SHG}}{P_{Pump}}$ from the measured spectrogram. It was observed from Fig.~\ref{second_harmonic_efficiency}b that the measured conversion efficiency (black dots) increases linearly with the increase of input pump power. Through  a linear fitting (red solid line in Fig.~\ref{second_harmonic_efficiency}b), a normalized conversion efficiency of 4.9\%/mW is determined. To the best of our knowledge, this SHG conversion efficiency in LN microdisk is higher than those published in early literatures~\cite{Wolf:18,PhysRevApplied.11.034026} and about half of the highest one achieved with the quasi-phase matching recently~\cite{PhysRevLett.122.173903}. The result indicates that the pentagon mode observed in the high-Q LN microdisk can provide promising potential for highly efficient nonlinear wavelength conversion. Besides the high optical Q and resulting large intracavity intensity establishment, we believe the high efficiency is due to the large modal overlap between the fundamental and second harmonic modes. Note that the modal overlap of conventional whispering gallery modes between the fundamental and second harmonic light is small due to the significant difference between the wavelengths. In the case of polygon mode, on the other hand, may reach higher overlap if both modes forms polygon of the same order.

\subsection*{Optomechanical Oscillation}
\begin{figure}[hbtp]
   \centering
  \includegraphics[width=0.90\textwidth]{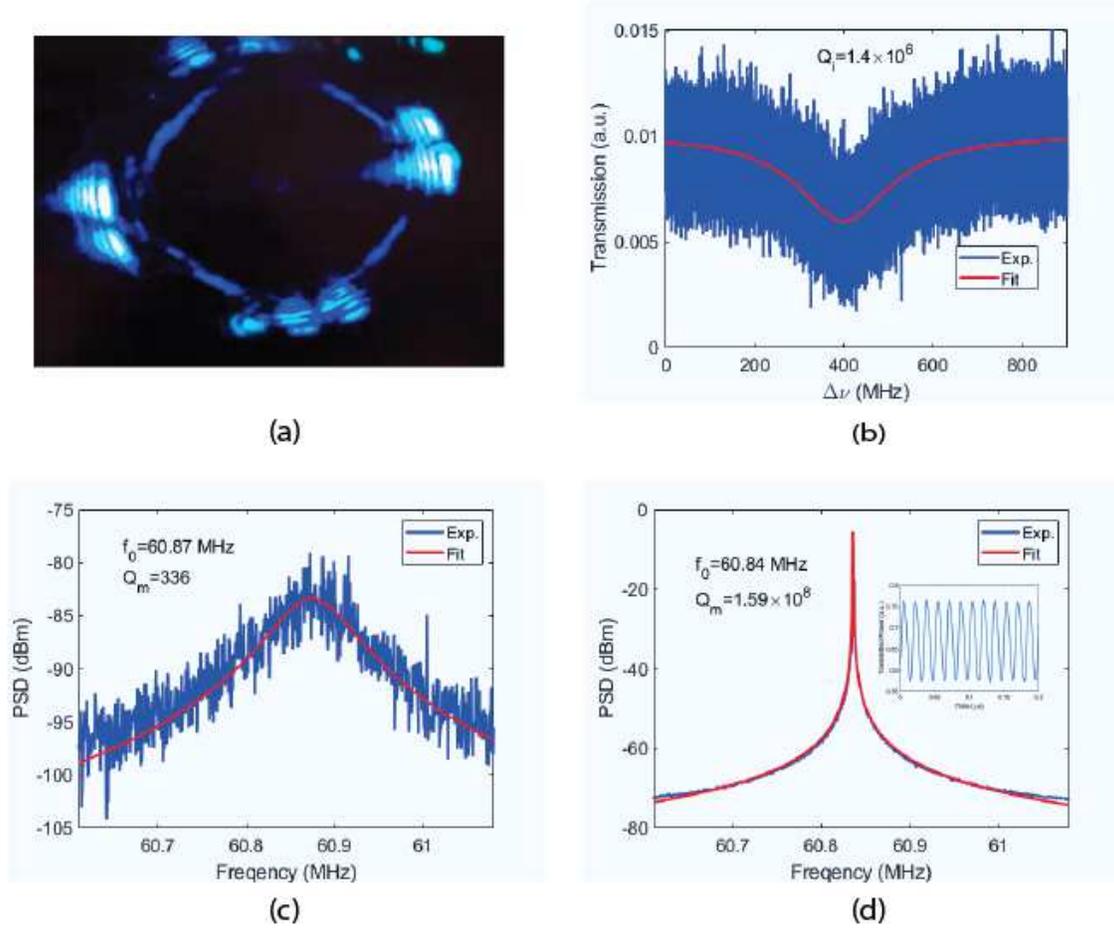}
   \caption{(a) Micrograph of a microdisk in square mode that triggers cavity optomechanical oscillation. (b) The mode displays an optical intrinsic Q of $1.4{\times}10^6$. (c) The intrinsic mechanical Q measured in air is 336. (d) When optomechanical oscillation in coherent regenerative regime, an effective mechanical Q as high as $1.59{\times}10^8$ is reached. Inset: oscilloscope trace of the transmitted spectrum vs. time.}\label{Fig_optomechanics}
\end{figure}

The high Q polygon mode further enables cavity optomechanics, for the first time according to the authors' knowledge. As shown in Fig.~\ref{Fig_optomechanics}a and ~\ref{Fig_optomechanics}b, a square mode with an intrinsic optical Q as high as $1.4{\times}10^6$ was observed. As a result, coherent regenerative optomechanical oscillation was observed using our 970~nm laser as a pump. The electrical spectrum of the transmitted signal shown in Fig.~\ref{Fig_optomechanics}c further demontrated that the oscillation processes an intrinsic mechanical quality factor $Q_m=336$ which is close to those excited by conventional whispering modes~\cite{Fang:19}. While in coherent regenerative regime, an effective mechanical quality factor as high as $Q_m=1.59\times{10^8}$ (the transmitted power vs. time is plotted as the inset), again close to the conventional whispering modes excitation. In contrast, cavity optical mechanics haven't been observed in the deformed cavities due to the combined difficulties in maintaining high optical and mechanical quality factors. It is worth emphasizing that the cavity optomechanical oscillation triggered by the optical forces only at the vertices of the polygon modes, which has significantly smaller opto/mechanical interaction area compared to the conventional whispering gallery counterpart. Such difference, may, at least in principle, excite mechanical oscillations localized at the vertices and reduce the damping for higher the mechanical quality factor.

\subsection*{Frequency comb generation}
\begin{figure}[hbtp]
   \centering
  \includegraphics[width=0.90\textwidth]{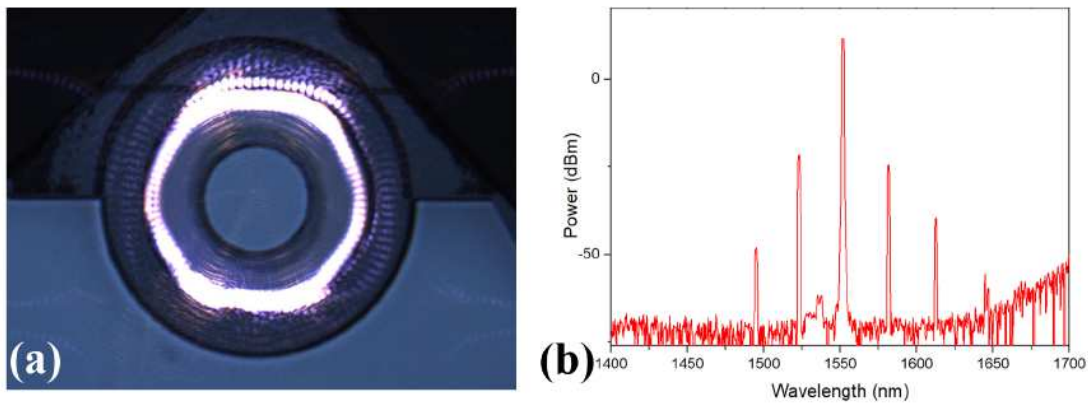}
   \caption{Experiment setup for measuring the SHG and optomechanics modes of the LN microdisk. The pump coupled in the LN microdisk with the fiber taper. The SHG signals scattered from the disk are focused by an objective lens before being collected by a spectrometer.}\label{Fig_frequencycomb}
\end{figure}

We further extend our experiment into frequency comb generation.  As shown in Fig.~\ref{Fig_frequencycomb}a, using a 1550~nm laser amplified by an optical amplifier, frequency comb was generated in hexagon mode.  It is worth noting that in this particular experiment, the hexagon vertices does not reach at the edge of the disk, indicating that dispersion engineering through modify the shape of the disk is unnecessary. Although the exact mechanism is still under investigation, we believe that formation of the polygon mode may inherently satisfy the dispersion relation required by comb generation, which may provide us a unique and efficient methods for comb generation.

\section*{Discussion}
To conclude, we demonstrate high Q polygon optical modes in the LN microdisk, which yields high efficiency in second harmonic, cavity optomechanics and frequency comb generations. In second harmonic generation, high efficiency of 4.9\%/mW was reached. The unique taper/cavity coupling mechanism provides a stable, robust and easy to align solution which is otherwise challenging in conventional whispering gallery mode coupling mechanisms. In the future research, the theoretical details will be laid out through full vector, full wave numerical simulations. Through such modelling efforts it maybe possible to achieve high Q optomechanical oscillation verify the dispersion self-adaptive nature in frequency comb generations. 

\section*{Methods}
\subsection{Device Fabrication and characterization} 
\begin{figure}[hbtp]
   \centering
  \includegraphics[width=0.90\textwidth]{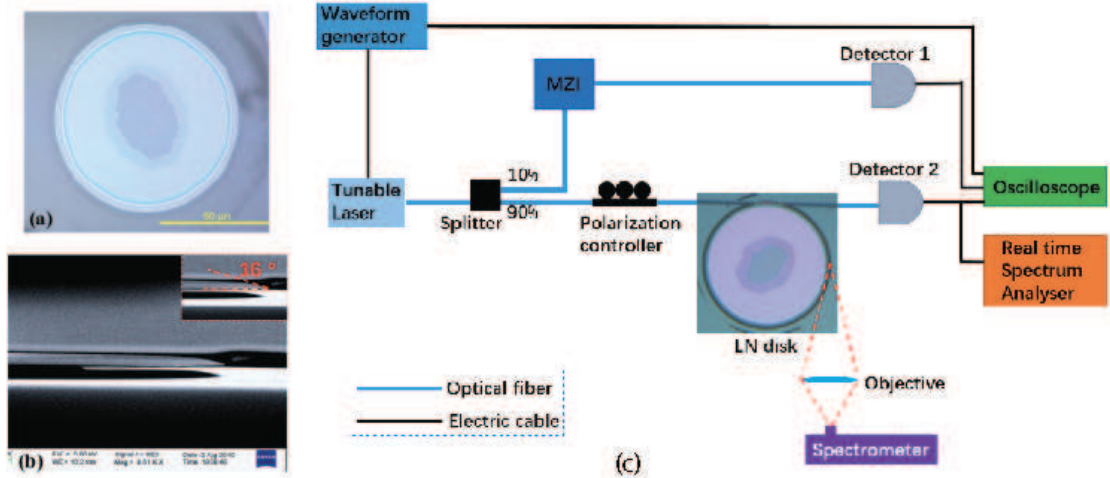}
   \caption{(a) Optical micrograph of a 84~$\mathrm{\mu}$m LN microdisk. (b) SEM micrograph shows the disk has a wedge angle of 16$^\circ$, (c)Experiment setup for measureing the SHG and optomechanics modes of the LN microdisk. The pump coupled in the LN microdisk with the fiber taper. The SHG signals scattered from the disk are focused by an objective lens before being collected by a spectrometer.}\label{Fig_sem}
\end{figure}

In this paper, we choose a high-Q lithium niobate microdisk as our platform. It is enabled by the rapid development of micro- and nanofabrication technologies for lithium niobate on insulator (LNOI)~\cite{Boes_Status,lin_fabrication_2015,lin_phase-matched_2016,fang_monolithic_2017,fang_fabrication_2017}. In particular, the very recent achievement on the realization of LNOI microresonators with ultrahigh Q factors above $10^7$ using a chemo-mechanical polishing lithography opens the venue to a broad spectrum of applications ranging from nonlinear optics and optomechanics to high-sensitivity sensing and precision metrology. The first results in the ultrahigh-Q microresonators have demonstrated rich nonlinear optical phenomena as well as high-Q mechanical modes at frequencies as high as 100.23~MHz~\cite{luo_chip_2017,wu_lithium_2018,Fang:19}, which aggressively motivates us to search for unconventional effects where optical Q need to be at its extreme.
In our experiment, the on-chip LN microdisk resonator was fabricated on a commercially available Z-cut LN thin film wafer with a thickness of 700~nm (NANOLN, Jinan Jingzheng Electronics Co., Ltd). The LN thin film is bonded by a silica layer with a thickness of ${\sim}\mathrm{2~{\mu}m}$ on a $\mathrm{500~{\mu}m}$ thick LN substrate. The LN microdisk is fabricated using space-selective femtosecond laser direct writing followed by chemo-mechanical polishing (CMP). More details about the device fabrication can be found in Ref.~\cite{Rongbo_Long,wang_chemo-mechanical_2019,Fang:20}. Fig.~\ref{Fig_sem}a and~\ref{Fig_sem}a shows the freestanding LN microdisk with a diameter of $\mathrm{84~\mu{m}}$ supported by a fused silica pillar with a diameter of $\mathrm{\sim~25~\mu{m}}$. The rim of the LN microdisk displays interference patterns under illumination, indicating the varying thickness at the edge of the LN disk. The wedge the wedge angle of 16$^\circ$ is observed. The extended wedge helps to enhance the optical Q factors as the light scattering from the edge tip is significantly reduced~\cite{Lee2012}.
\subsection{Experiment setup}
The experiment setup is shown in Fig.~\ref{Fig_sem}c. In our experiments, continuous-wave tunable diode lasers at various wavelengths were used. We can linearly scanned the optical frequency of the tunable diode laser using a waveform generator. The 90\% port of the fiber optic splitter was connected to the input of a tapered fiber for coupling the pump light into the LN microdisk. The tapered fiber was fabricated by pulling a section of single mode fiber (HI~1030) to a diameter of $\mathrm{1~{\mu}m}$. The polarization states were selectively excited using an in-line fiber polarization controller. A photoreceiver (New focus~1811) was connected to the output of the tapered fiber for capturing the optical signal. The captured signal was then converted to electrical signals and analyzed by an oscilloscope (Agilent~DSO90404A) and a real-time spectrum analyzer (Tektronix RSA~3408B). To measure the optical Q of a cavity resonant mode, the 10\% output branch of the fiber optic splitter was connected to an optical fiber Mach-Zehnder interferometer (MZI) thermally stabilized in a foam box filled with ice water mixture. The transmitted optical signal from the MZI was captured by the oscilloscope synchronous with the signal out from the tapered fiber. Meanwhile, a 20$\times$ objective lens was used to collect the scattering light from the edge of the LN microdisk. In the experiment of the second harmonic generation, the spectrum of scattered fundamental and second harmonic lights were measured by a spectrometer (Ocean Optics~USB4000). We do not use the tapered fiber to collect the second harmonic signal, because the short wavelength at 485~nm is difficult to be coupled out from LN disk to the tapered fiber fabricated.
\section*{Data availability} The data that support the findings of this study are available from the corresponding authors on request.

\noindent\textbf{Acknowledgements}Z.F. J.L., R.W., J.Z.,Z.W., M.W., Y.C.acknowledge the support fromNational Key R\&DProgram of China (2019YFA0705000, 2018YFB2200400),National Natural Science Foundation of China (11822410, 11874154, 11874375, 11734009, 61761136006, 11674340, 61675220, 61590934), the Strategic Priority Research Program of CAS (XDB16030300), the Key Project of the Shanghai Science and Technology Committee (18DZ1112700, 17JC1400400), Shanghai Municipal Science and Technology Major Project (2019SHZDZX01), the Shanghai Pujiang Program (18PJ1403300), and Key Research Program of Frontier Sciences, CAS (QYZDJ-SSWSLH010). S.H.,S.F.,H.L. and T.L. acknowledge the support from Nature Science and Engineering Research Council of Canada (NSERC) Discovery (Grant No. RGPIN-2020-05938), CFI LOF and Threat Reduction Agency (DTRA) Thrust Area 7, Topic G18 (Grant No. GRANT12500317).

\noindent\textbf{Author Contributions} Z.F. fabricated the devices, prepared the samples. Z.F. and S.H. performed the experiments and processed the data. F.S. and T.L. performed numerical simulations. Y. C. and T. L. conceived the concept and supervised the project. Z.F., S.H., F.S., H.L., J.L., R.W., J.Z.,Z.W., M.W., Y.C., and T.L. worked together on the result analysis/interpretation and manuscript preparation.

\noindent\textbf{Author Information} The authors declare no competing financial interests.  Correspondence and requests for materials should be addressed to Y.C. ($ya.cheng@siom.ac.cn$) and T.L. ($taolu@ece.uvic.ca$) .

\end{document}